\documentclass[iop]{emulateapj}

\newcommand{\ts}{\thinspace}
\newcommand{\etal}{\mbox{et\ts al.\ts}}

\newcommand{\OIII}{[O\,{\footnotesize III}]\,}
\newcommand{\NII}{[N\,{\footnotesize II}]}

\newcommand{\microns}{$\mu$m }

\shorttitle{ }
\shortauthors{Kartaltepe \etal}

\begin{document}

\title{Rest-frame Optical Emission Lines in Far-Infrared Selected Galaxies at $z<1.7$ from the FMOS-COSMOS Survey}

\author{Jeyhan S. Kartaltepe\altaffilmark{1,2}, D. B. Sanders\altaffilmark{3}, J. D. Silverman\altaffilmark{4}, D. Kashino\altaffilmark{5}, J. Chu\altaffilmark{3}, H. Zahid\altaffilmark{3}, G. Hasinger\altaffilmark{3}, L. Kewley\altaffilmark{6}, K. Matsuoka\altaffilmark{7},  T. Nagao\altaffilmark{8}, L. Riguccini\altaffilmark{9},  M. Salvato\altaffilmark{10,11}, K. Schawinski\altaffilmark{12}, Y. Taniguchi\altaffilmark{8}, E. Treister\altaffilmark{13}, P. Capak\altaffilmark{14}, E. Daddi\altaffilmark{15}, K. Ohta\altaffilmark{7}
}

\altaffiltext{1}{National Optical Astronomy Observatory, 950 N. Cherry Ave., Tucson, AZ, 85719, email: jeyhan@noao.edu}
\altaffiltext{2}{Hubble Fellow}

\altaffiltext{3}{Institute for Astronomy, 2680 Woodlawn Dr., University of Hawaii, Honolulu, HI, 96822}

\altaffiltext{4}{Kavli Institute for the Physics and Mathematics of the Universe, Todai Institutes for Advanced Study, the University of Tokyo, Kashiwa, Japan 277-8583 (Kavli IPMU, WPI)}

\altaffiltext{5}{Division of Particle and Astrophysical Science, Graduate School of Science, Nagoya University, Nagoya 464-8602, Japan}

\altaffiltext{6}{Research School of Astronomy and Astrophysics, The Australian National University, Cotter Road, Weston Creek, ACT 2611}

\altaffiltext{7}{Department of Astronomy, Kyoto University, Kitashirakawa-Oiwake-cho, Sakyo-ku, Kyoto 606-8502, Japan}

\altaffiltext{8}{Research Center for Space and Cosmic Evolution, Ehime University, Bunkyo-cho 2-5, Matsuyama, Ehime 790-8577, Japan}

\altaffiltext{9}{Observat—rio do Valongo, Universidade Federal do Rio de Janeiro, Ladeira Pedro Antonio, 43, Saœde, Rio de JaneiroÐRJ, CEP 22240-060}

\altaffiltext{10}{Max Planck Institut fŸr Plasma Physik, Giessenbachstrasse 1, 85748 Garching, Germany}

\altaffiltext{11}{Excellence Cluster, Boltzmann Strasse 2m 85748, Garching, Germany}

\altaffiltext{12}{Institute for Astronomy, Department of Physics, ETH Zurich, Wolfgang-Pauli-Strasse 27, CH-8093 Zurich, Switzerland}

\altaffiltext{13}{Universidad de Concepci\'on, Departamento de Astronom\'ia, Casilla 160-C, Concepci\'on, Chile}

\altaffiltext{14}{Spitzer Science Center, California Institute of Technology, Pasadena, CA 91125, USA }

\altaffiltext{15}{CNRS, AIM-Unit«e Mixte de Recherche CEA-CNRS-Universit«e Paris VII-UMR 7158, F-91191 Gif-sur-Yvette, France. }


\begin{abstract}

We have used FMOS on Subaru to obtain near-infrared spectroscopy of 123 far-infrared selected galaxies in COSMOS and obtain the key rest-frame optical emission lines. This is the largest sample of infrared galaxies with near-infrared spectroscopy at these redshifts. The far-infrared selection results in a sample of galaxies that are massive systems that span a range of metallicities in comparison with previous optically selected surveys, and thus has a higher AGN fraction and better samples the AGN branch. We establish the presence of AGN and starbursts in this sample of (U)LIRGs selected as {\it Herschel}-PACS and {\it Spitzer}-MIPS detections in two redshift bins ($z\sim0.7$ and $z\sim1.5$) and test the redshift dependence of diagnostics used to separate AGN from star-formation dominated galaxies. In addition, we construct a low redshift ($z\sim0.1$) comparison sample of infrared selected galaxies and find that the evolution from $z\sim1.5$ to today is consistent with an evolving AGN selection line and  a range of ISM conditions and metallicities from the models of  \cite{2013ApJ...774L..10K}. We find that a large fraction of (U)LIRGs are BPT-selected AGN using their new, redshift-dependent classification line. We compare the position of known X-ray detected AGN (67 in total) with the BPT selection and find that the new classification line accurately selects most of these objects ($>70\%$). Furthermore, we identify 35 new (likely obscured) AGN not selected as such by their X-ray emission. Our results have direct implications for AGN selection at higher redshift with either current (MOSFIRE, KMOS) or future (PFS, MOONS) spectroscopic efforts with near-infrared spectral coverage.

\end{abstract}



\section{Introduction}

Not long after they were initially discovered, it was shown that the high infrared luminosities ($L_{\rm IR}$) of luminous and ultraluminous infrared galaxies (LIRGs: $L_{\rm IR} >10^{11} L_{\odot}$, ULIRGs: $L_{\rm IR} >10^{12} L_{\odot}$, (U)LIRGs collectively) originate from extreme star formation, AGN activity, or a combination of the two  (see review by Sanders \& Mirabel 1996). In the local universe, both the merger fraction \cite[e.g.,][]{2002ApJS..143..315V} and AGN fraction \cite[e.g.,][]{1995ApJS...98..171V,Tran:2001p3271} increases systematically with $L_{\rm IR}$. These two observational results support the `merger scenario' initially proposed by \cite{Sanders:1996p1630} where (U)LIRGs represent a transition stage between gas rich spiral galaxies and red elllipticals. Galaxies enter this transition stage through a major merger \cite[e.g.,][]{Toomre:1972p4262}, which triggers star formation and fuels a central black hole. During this stage, the black hole is enshrouded by dust and later, once star formation begins to subside, the remnant evolves into an optical QSO.  Thus, understanding the relationship between star formation and AGN activity, and how each contribute to the total $L_{\rm IR}$ of galaxies, is a critical test of the merger scenario. While many such studies of (U)LIRGs in the local universe have been conducted \cite[e.g.,][]{2010ApJ...709..884Y}, the relative role of these two processes at high redshift ($z>0.5$), where these objects dominate cosmic star formation activity, have yet to be thoroughly explored.

Previous studies of high redshift (U)LIRGs have found that while the merger fraction also systematically increases with $L_{\rm IR}$ \cite[e.g.,][]{2010ApJ...721...98K, 2012ApJ...757...23K}, the absolute fraction that have clearly gone through a major merger is lower than in the local universe, suggesting that such an event may not be necessary for these extreme luminosities at high redshift. However, the difficulty of identifying merger signatures at high redshift means that these fractions should only be considered lower limits \cite[e.g.,][]{2014ApJ...791...63H}. On the other hand, the AGN fraction among high redshift (U)LIRGs is similar to that of their local counterparts \cite[e.g.,][]{2010ApJ...709..572K,2013MNRAS.433.1015S,2013ApJ...764..176J}. Identifying AGN among these dust enshrouded objects, some of which are likely to be Compton thick, can be difficult. Various AGN selection techniques have been used in the past, such as selecting objects with high X-ray luminosities, or those with power-law slopes in the mid-infrared \cite[e.g.,][]{Donley:2007p3092}. One of the classic AGN identification techniques is through nebular emission line diagnostics \cite[e.g.,][]{1981PASP...93....5B,1987ApJS...63..295V}. Since each of these methods is sensitive to AGN with different redshifts, luminosities, and dust properties, in order to obtain a full AGN census among a sample of galaxies it is essential to combine multiple selection techniques \cite[e.g.,][]{Hickox:2009p4151}.

We use the `BPT diagram' \citep{1981PASP...93....5B} to identify AGN among high redshift (U)LIRGs selected from observations taken with the {\it Herschel Space Observatory} and {\it Spitzer Space Telescope} of the COSMOS field \citep{Scoville:2007p1776} and compare them to known AGN identified in the X-ray. Since the BPT emission lines are shifted out of the optical at $z\sim0.5$ and $z\sim1.0$ for the two sets of lines, respectively, near-infrared spectroscopy is essential for obtaining these lines at higher redshifts. Until recently, this was only possible for small numbers of objects using long slit spectroscopy, but now that several multi-object NIR spectrographs are available, larger surveys are possible. In order to apply the BPT diagnostic at high redshift, it is important to understand how the AGN selection lines \citep{2001ApJ...556..121K,Kewley:2006p247,2003MNRAS.346.1055K} evolve with redshift. Using theoretical models, \cite{2013ApJ...774..100K,2013ApJ...774L..10K} have predicted that the BPT line ratios should change as a function of redshift due to the evolving ISM conditions and derived a new classification line to separate AGN from star forming galaxies. The first large surveys of star-forming galaxies in the near-infrared (e.g., \citealt{2012PASJ...64...60Y,2014MNRAS.437.3647Y} as shown by \citealt{2013ApJ...774L..10K} and \citealt{2014ApJ...788...88J}; \citealt{2014ApJ...792...75Z,2014ApJ...795..165S}; \citealt{2014arXiv1409.6522C}) find that the observed line ratios evolve in the way predicted by Kewley et al., but the AGN samples in these studies are small and insufficient to test these diagnostics.

Here, we present results from the low-resolution FMOS \citep{2010PASJ...62.1135K} survey of the COSMOS field (Kartaltepe et al., in prep), and use the BPT diagnostic to identify AGN among high redshift (U)LIRGs for 59 galaxies at $z\sim 0.7$ and 64 galaxies at $z\sim1.5$. This Letter is organized as follows: In Sections 2 and 3, we present the FMOS observations and ancillary datasets used in our analysis, respectively. We present the BPT diagram in Section 4, compare to previous results in Section 5, and summarize our results in Section 6.

\section{FMOS Observations}

The wide field-of-view of FMOS ($30\arcmin$) and the large number of fibers ($\sim400$) make it ideal for a wide-area survey over the 2-deg$^{2}$ COSMOS field. The FMOS survey of the COSMOS field is divided into two parts -- now complete low-resolution survey ($R\sim 600$; Kartaltepe et al., in prep) and an ongoing high-resolution survey ($R\sim 2000$; \citealt{2014arXiv1409.0447S}). Here, we focus on observations from the low-resolution survey since all four of the BPT diagnostic emission lines are obtained at once for galaxies at $z\sim 1.5$, the same redshift range where most of the {\it Herschel} detected ULIRGs lie, as highlighted in Figure 1. In the low-resolution mode, FMOS covers the wavelength range $0.9-1.8$ \microns with a dispersion of $\sim 5 \AA$ per pixel. With this wavelength coverage, H$\alpha$ and \NII\ can be observed at $0.5<z<1.7$ and \OIII and H$\beta$ at $1.0<z<2.7$. All four lines can be observed at $1.0<z<1.7$.

\begin{figure*}
\plotone{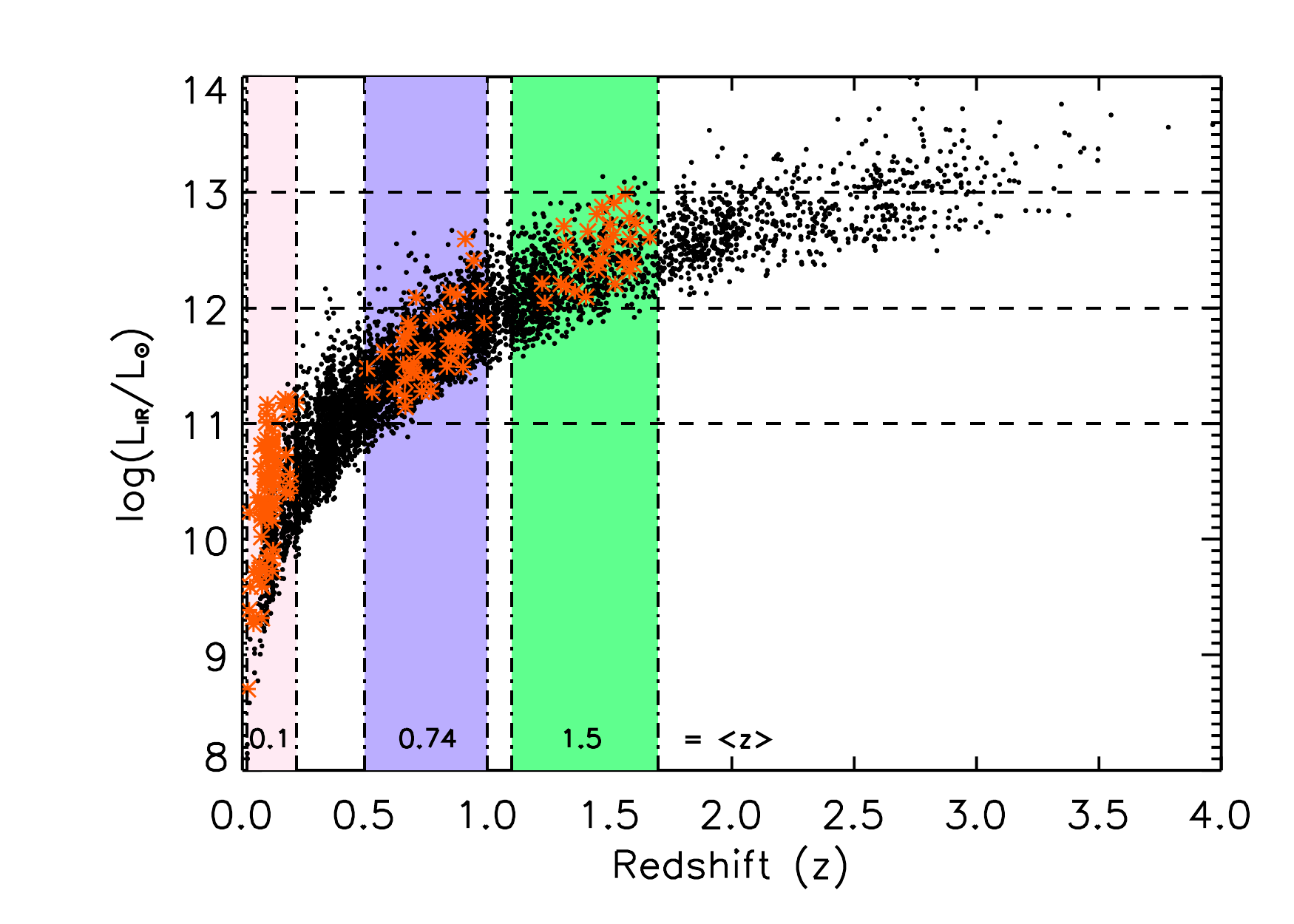}
\caption{Total infrared luminosity ($L_{\rm IR}$) as a function of redshift for the {\it Herschel}-PACS sample. Horizontal dashed lines mark the LIRG, ULIRG, and HyLIRG luminosity divides. The colored vertical bands highlight our three redshift bins. The low-redshift bin ($<z>=0.1$) contains galaxies with all four lines (\NII, H$\alpha$, \OIII, and H$\beta$) in the optical, the intermediate-redshift bin ($<z>=0.7$) has \NII\ and H$\alpha$ in the near-infrared and \OIII\ and H$\beta$ in the optical, and the high-redshift bin ($<z>=1.5$) has all four lines in the near-infrared. Over-plotted in orange are the galaxies with measured emission lines presented in this paper.} 
\label{lir}
\end{figure*}

We observed 19 pointings over the entire COSMOS field, with typical integration times of 180 minutes per pointing in varying weather conditions. We observed in cross-beam switching mode, with two fibers assigned to each target -- one for the target and the other for sky -- and dithered between the two for optimum sky subtraction. The observations were reduced, wavelength, and flux-calibrated using the publicly available pipeline FIBRE-pac (FMOS Image-Based REduction package; \citealt{2012PASJ...64...59I}). We visually inspected the 1D and 2D reduced spectra from all 19 pointings and measured redshifts using SPECpro \citep{2011PASP..123..638M}. Each spectrum was inspected by two of the authors and the results were compared. In cases of discrepant redshift measurements, we inspected the spectrum again and chose a final redshift. In total, we measured redshifts and quality flags for 988 objects. The target selection for the full survey will be discussed in a future paper. Here we focus on the key targets for the survey -- far-infrared selected galaxies. All of the objects discussed in this paper have a redshift quality flag of 4 since we require the detection of at least three of the key emission lines.

\section{Ancillary datasets}

In addition to the FMOS observations, we use optical spectroscopic observations from several other sources, including VIMOS (Visible Multi-Object Spectrograph; Le F'evre et al. 2003) observations from zCOSMOS \citep{Lilly:2007p2297}, DEIMOS observations from Keck II \citep{2010ApJ...709..572K}, and the Sloan Digital Sky Survey (SDSS; \citealt{2009ApJS..182..543A}). These optical spectra were used for all of the diagnostic lines in the low redshift comparison sample and for H$\beta$ and \OIII in the intermediate redshift sample.

The sample of infrared galaxies in this paper were selected using {\it Herschel} observations of the COSMOS field from the PEP (PACS Evolutionary Probe; \citealt{2011A&A...532A..90L}) and HerMES (Herschel Multi-tiered Extragalactic Survey; \citealt{2012MNRAS.424.1614O}) surveys using the PACS ($100-160$\ts$\mu$m) and SPIRE ($250-500$\ts$\mu$m) instruments, respectively. In addition, we selected targets using the {\it Spitzer}-MIPS $24\ts\mu$m observations of COSMOS from S-COSMOS \citep{2007ApJS..172...86S}. The detections, source photometry, and counterpart matching (using $24\ts\mu$m priors) is fully described in \cite{2013ApJ...778..131L}.

The $L_{\rm IR}$ for the entire {\it Herschel}-PACS sample is shown in Figure 1 as a function of redshift. We derived $L_{\rm IR}$ for each source by fitting the MIR-FIR SED (using the IRAC 8\ts$\mu$m, MIPS 24\ts$\mu$m, 70\ts$\mu$m, PACS 100\ts$\mu$m and 160\ts$\mu$m, and SPIRE 250\ts$\mu$m photometry) with several template libraries \citep{Chary:2001p2083, Dale:2002p2130, Lagache:2003p1825, Siebenmorgen:2007p2697} using the SED fitting code {\it Le Phare}\footnote{http$://$www.cfht.hawaii.edu/$\sim$arnouts/LEPHARE/cfht$\_$lephare/lephare.html} written by S. Arnouts and O. Ilbert. The best fit model was chosen by finding the one with the lowest $\chi^{2}$ value and allowing for rescaling of the templates and $L_{\rm IR}$ was then calculated from the best fit template by integrating from 8--1000\ts${\mu}$m.

We used the ancillary data to identify AGN among our FIR sample. We used the catalogs of \cite{2010ApJ...716..348B} and \cite{2012ApJS..201...30C}, which identify sources associated with XMM-Newton \citep{Hasinger:2007p2291} and Chandra \citep{2009ApJS..184..158E} detections. We also identified IR-selected AGN candidates based on their IRAC colors following \citealt{2012ApJ...748..142D}. We choose this selection method because it is the most conservative at excluding star-forming contaminants.

\section{BPT Diagram }

\begin{figure*}
\plotone{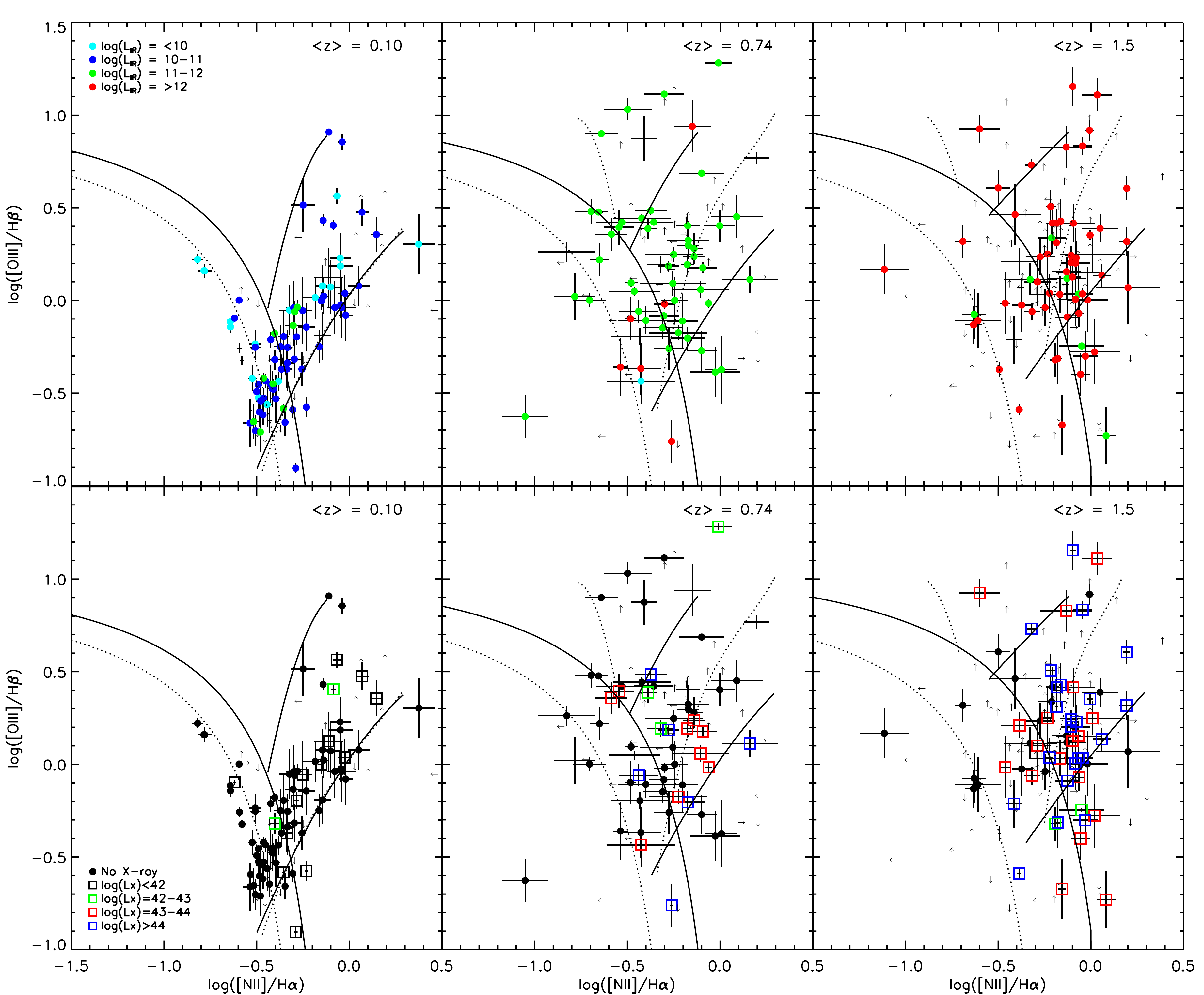}
\caption{The BPT diagram, \OIII$/$H$\beta$ versus \NII$/$H$\alpha$, for galaxies in each of the three redshift bins in our sample with the lower-limit abundance sequence (dotted curve), the redshift-dependent AGN classification line (solid curve) and the mixing sequences for Scenarios 3 and 4 (solid and dotted lines, respectively; Kewley \etal 2013a,b) over-plotted. 1$\sigma$ uncertainties are shown for each galaxy. Note that the dispersion in the location of the galaxies on the diagram is larger than the typical error bar. The points in the top row are color-coded by their total infrared luminosity while those in the bottom row are marked with a box if they are detected in the X-ray and color-coded by their X-ray luminosities.} 
\label{BPT}
\end{figure*}

We measured emission line fluxes for all of the galaxies in the FMOS sample by fitting a gaussian to each line using a custom script allowing the line widths and fluxes to be free parameters. We also fit lines from the zCOSMOS and DEIMOS spectra for galaxies in the redshift range where H$\alpha$ and \NII\ are observed by FMOS but H$\beta$ and \OIII fall in the optical. We corrected for photospheric Balmer absorption following the method of Zahid et al. 2014. In addition, we used these spectra to select a low redshift sample of galaxies of  {\it Herschel}-detected infrared galaxies, where all four lines are observed at optical wavelengths, including publicly available SDSS spectra from DR7. For the SDSS galaxies, we used line flux measurements from the MPA/JHU DR7 release.\footnote{http$://$www.mpa-garching.mpg.de/SDSS/DR7/} For our final sample, we selected all galaxies with $S/N>3$ (corresponding to a flux limit of $\sim9\times 10^{-17}$ erg cm$^{-2}$ s$^{-1}$ for FMOS -- the flux limit for the optical spectra is lower and different for each survey) in all four of the diagnostic emission lines, excluding broad line AGN. The three redshift bins for our sample are highlighted in Figure 1 and summarized in Table 1. The low redshift bin is almost entirely made up of sources with $L_{\rm IR} < 10^{11}\ts L_{\odot}$ (83 galaxies in total), the intermediate redshift of mostly LIRGs and low luminosity ULIRGs (59 galaxies), and the high redshift bin of mostly ULIRGs with some high luminosity LIRGs (64 galaxies). The median stellar masses for this sample are $<log(M/M_{\odot})>=10.4$, 10.2, and 10.4 for each respective redshift bin. In addition, there are 39 and 65 galaxies in the intermediate and high redshift bins, respectively, with $S/N>3$ detections in three out of four of the emission lines, enabling us to place a lower or upper limit on the BPT diagram for these galaxies. This represents the largest sample of high redshift (U)LIRGs with measurements of these key diagnostic lines in the literature -- the properties of this sample are given in Table 2.

\begin{figure*}
\plotone{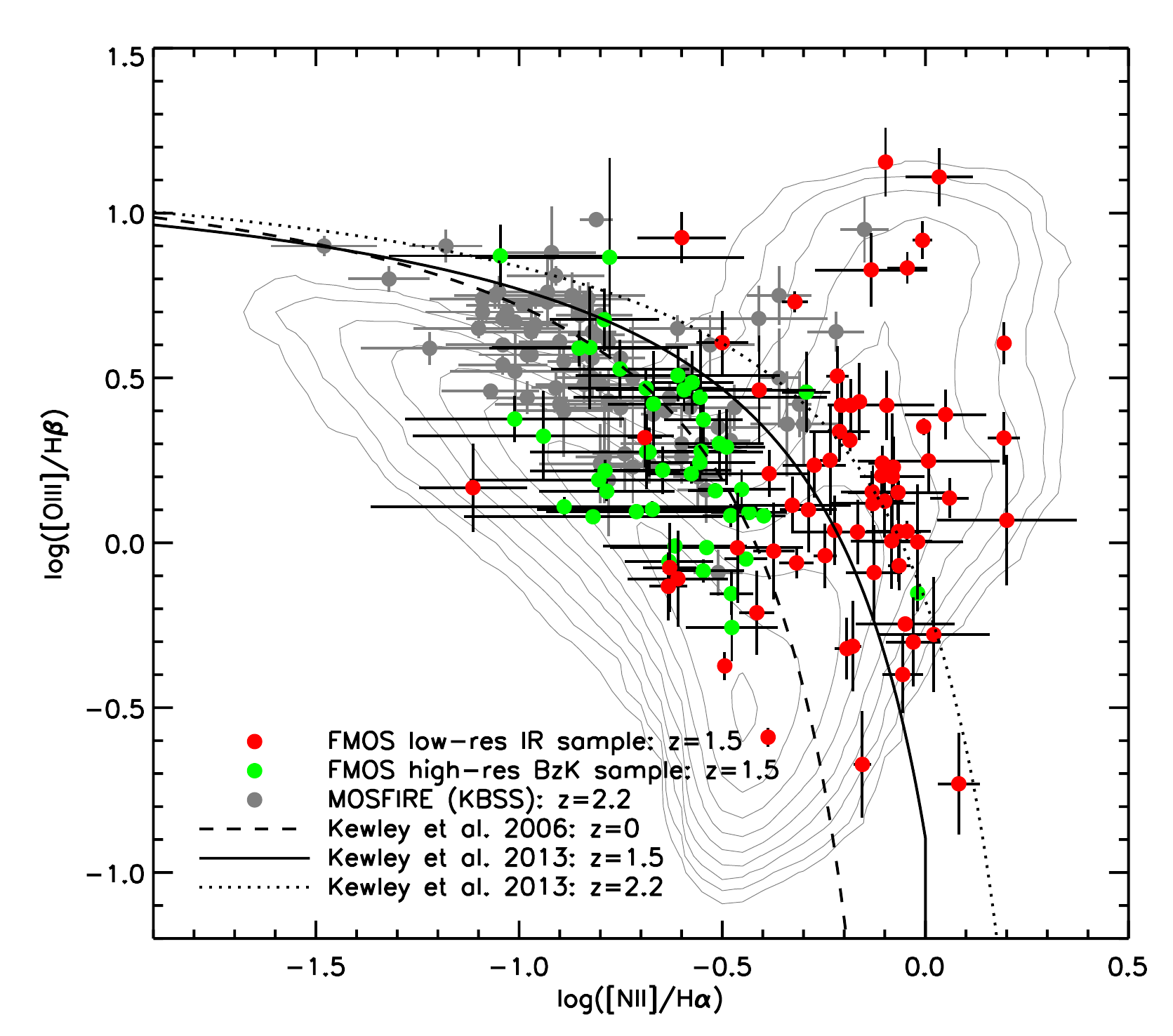}
\caption{The BPT diagram, \OIII$/$H$\beta$ versus \NII$/$H$\alpha$, for two optically selected galaxy samples at high redshift with near-infrared spectroscopy at $z\sim1.5$ (green, from the high resolution FMOS program; \citealt{2014arXiv1409.0447S}) and $z\sim 2.2$ (gray, from MOSFIRE; \citealt{2014ApJ...795..165S}) in comparison to our infrared selected sample at $z\sim 1.5$ (red). Over-plotted are the AGN classification lines at $z=0$, $z=1.5$, and $z=2.2$ from Kewley \etal (2006, 2013) and the contours of the full SDSS sample at low redshift. Note the offset between the infrared and optically selected samples.}
\label{sf}
\end{figure*}

The BPT diagram is shown in Figure 2 for all three redshift bins. The top panel shows the galaxies color-coded by their infrared luminosity and the bottom panel highlights whether a given galaxy is also identified as an AGN in the X-ray and color-coded by their X-ray luminosity. The luminosity limit is redshift dependent (log$(L_{\rm X})\sim42.5$ erg\ts s$^{-1}$ at $z\sim0.7$ and log$(L_{\rm X})\sim43.3$ erg\ts s$^{-1}$ at $z\sim1.5$ for the 2-10 keV band for the Chandra observations). Essentially all of X-ray detections at $z>0.5$, have  $L_{\rm X}>10^{42} erg\ts s^{-1}$, a luminosity level typically due to an AGN and a higher X-ray luminosity than would be expected from star formation using the relation of Mineo et al. 2014. It should be noted that our observing program specifically targeted X-ray AGN whenever possible so the total fraction of X-ray AGN among our (U)LIRGs is high. Therefore, the total AGN fractions quoted here should not be considered as absolutes, but rather the comparison between AGN identified via the different methods is what we are interested in. For comparison, $\sim 15-30\%$ of (U)LIRGs at this redshift in COSMOS are X-ray detected AGN \cite{2010ApJ...709..572K}. Over-plotted on each panel are the lower-limit abundance sequence, the redshift-dependent AGN classification line, and the Starburst-AGN mixing sequence for Scenarios 3 and 4 from \cite{2013ApJ...774L..10K}. We identify all galaxies above the AGN classification line as `BPT-selected AGN,' although this new dividing line is uncertain and likely does not select all AGN (especially those in composite systems). The mixing sequences range from Scenarios 1-4 and span both normal and extreme ISM conditions and metal-rich and metal-poor AGN narrow line regions (NLR) at high redshift. Here, we plot scenario 3 (extreme ISM conditions and metal-rich AGN NLR) and scenario 4 (extreme ISM conditions and metal-poor AGN NLR) since they appear to be the best match for our high redshift data points.  The percentage of sources that fall within the bounds for scenarios 1-4 are 62\%, 52\%, 66\%, and 64\% for the intermediate redshift and 70\%, 28\%, 73\%, and 53\% at high redshift. Scenario 2 (normal ISM conditions and metal-poor AGN NLR) has the lowest fraction, while the fraction in the other three scenarios is comparable.

Figure 2 and Table 1 indicate that a large fraction of the (U)LIRGs in our sample are BPT-selected AGN, using this new classification line, including many that only have upper limits for one of the lines. In the intermediate redshift bin, 35/47 of the LIRGs and 2/6 of the ULIRGs are BPT AGN. For the high redshift bin, these numbers are 5/7 LIRGs and 38/53 ULIRGs. The differences between the LIRGs and ULIRGs in these two bins are not likely to be significant since the detection limit of the {\it Herschel} data means that there is not a large dynamic range in the luminosities probed at a given redshift. For the intermediate redshift bin, 17/23 of the X-ray detected AGN are identified as BPT AGN and 34/44 in the high redshift bin. While these fractions are high, not every X-ray AGN is detected as an AGN on the BPT diagram. The remaining objects are likely to be either be low-luminosity AGN or composite objects -- a possible indication that it is more difficult to obtain a clean AGN selection at high redshift. A similar comparison to IR selected AGN candidates \citep{2012ApJ...748..142D} confirms the presence of an obscured AGN in all of the objects in the intermediate redshift bin and in 26/33 in the high redshift bin.

In total, there are 36 and 20 galaxies in the intermediate and high redshift bins that were not known to be AGN through an X-ray detection. Interestingly, 23 ($64\%$) of these galaxies in the intermediate and 12 ($60\%$) in the high redshift bin are detected as BPT AGN. Though our X-ray data are not deep enough (Chandra reaches $5.7\times 10^{-16}$ erg cm$^{-2}$ s$^{-1}$) to detect all unobscured AGN at these redshifts, it is possible that these galaxies are highly obscured and candidates for Compton Thick AGN. It is also possible that these galaxies are unobscured, low luminosity AGN, but the high infrared luminosities and the presence of power-law slopes in the mid-infrared (33\% in the high redshift bin), make this possibility less likely.

\section{Comparison with Other Surveys}

Figure 3 shows the BPT diagram for two large samples of star-forming galaxies with near-infrared spectroscopy. The first sample comes from the FMOS high resolution survey \citep{2014arXiv1409.0447S} and we have excluded known AGN and {\it Herschel} detected objects. These 44 objects have $z\sim 1.5$ and all but four of them fall below the \citet{2013ApJ...774L..10K} AGN classification line at that redshift. Also shown are the 82 objects from the MOSFIRE-KBSS sample from \cite{2014ApJ...795..165S} (their Table 1) selected to be star-forming in a variety of ways, but at higher redshift than the FMOS sample ($z\sim2.2$). Most of these objects also fall below the \citet{2013ApJ...774L..10K} classification line, though there a few that lie above. These optically and sBzK selected samples confirm the evolution of the star-forming locus of galaxies with redshift but they are in stark contrast to the infrared selected sample, the focus of this paper, over-plotted in red. The infrared sample has a much higher obscured AGN fraction and therefore tends to have more extreme ratios on the BPT diagram, at least partly due to the fact that the infrared sample selects galaxies with higher masses and specific star formation rates than the optically selected samples ($<log(M/M_{\odot})>=10.4$ versus 10.0). The location of known X-ray selected AGN on the BPT diagram at high redshift has been explored by \cite{2013ApJ...763L...6T} and \cite{2014arXiv1409.6522C}. These small samples show that most of these AGN lie above the \citet{2013ApJ...774L..10K} selection line, but a few lie in the \citet{Kewley:2006p247} AGN-star formation composite region, suggesting that the line ratios of AGN may not evolve as much as the Kewley et al. lines indicate. Our larger sample of X-ray AGN among our infrared galaxies shows that a large fraction of objects lies above the AGN classification line, but that some would also be considered composite sources following the local criteria.

To reiterate, an evolution of the dividing line between star-forming galaxies and AGN is supported by the data and by the changing ionization properties of star-forming galaxies as indicated by our FMOS sBzK sample (Silverman et al. 2014) and those from the KBSS survey (Steidel et al. 2014). Since it is difficult to cleanly separate AGN dominated sources from AGN-star forming composite sources in our infrared sample, the evolution of the locus of AGN is unclear and further work with large samples of known AGN-dominated sources is required. This likely adds to the challenges of cleanly identifying AGN at high redshift with optical and near-infrared spectroscopic surveys such as those being undertaken with MOSFIRE, KMOS and the future PFS survey.

\section{Summary}

We have presented results from the large low-resolution near-infrared FMOS near-infrared spectroscopic survey of the COSMOS field. This survey has enabled us to compile the largest sample of infrared-selected galaxies with all four key diagnostic lines at high redshift to date. From the analysis of these sources, we summarize our results as follows.

\begin{enumerate}
\item Our final sample contains infrared galaxies with emission line measurements of all four key diagnostic lines in two redshift bins -- an intermediate redshift bin ($z\sim0.74$) containing 59 galaxies and a high redshift bin ($z\sim1.5$) containing 64 galaxies. In addition, there are 39 and 65  galaxies in each of these bins with detections for three out of four lines. 

\item We present the BPT diagram for this sample and find a high fraction of BPT-selected AGN among (U)LIRGs in both redshift bins using the new redshift-dependent Kewley et al. 2013 classification scheme. Many of these objects were not previously known to be AGN, suggesting that NIR spectroscopy is essential for a complete census of AGN at these redshifts.

\item A high fraction ($>70\%$) of the X-ray detected objects in our sample are selected as AGN using the new classification scheme. The remaining sources are likely to star-forming/AGN composite sources, rather than low-luminosity AGN. 

\item  The line ratios of our sample of (U)LIRGs show a high level of agreement with the Scenario 1, 3, and 4 mixing sequences of \citet{2013ApJ...774L..10K} suggesting that these galaxies likely span a range of ISM conditions and metalicities.

\item In comparison to optically and sBzK selected samples, our far-infrared selected galaxies mostly lie at higher values of   \NII$/$H$\alpha$ and span a wide range in \OIII$/$H$\beta$. Therefore, our FIR sample appears to be dominated by AGN, have higher metallicities, and higher stellar masses.

\end{enumerate}

 \acknowledgments
 


\bibliographystyle{apj1}

\begin{deluxetable}{lccccccccc}
  \tablewidth{0pt}
  \tabletypesize{\scriptsize}
  \tablecaption{IR-selected FMOS Source Properties}
  \setlength{\tabcolsep}{0.05in}
   \tablenum{2}
\tablehead{\colhead{ID}  & \colhead{RA} & \colhead{Dec} & 
              \colhead{Redshift} & \colhead{log($L_{\rm IR}/L_{\odot})$} & \colhead{log($L_{\rm X}$)} & 
              \colhead{\NII$/$H$\alpha$} & \colhead{\OIII$/$H$\beta$} & \colhead{IR AGN} & \colhead{BPT AGN} 
              }
\startdata
COSMOS J100057.20+020322.30 &   150.23833  &  2.05619 & 1.506 & 12.3 & 43.1 & 0.034 & 1.109 & N & Y \\
COSMOS J100023.01+020842.60 &   150.09587  &  2.14516 & 1.327 & 11.8 & 43.9 & 0.082 & -0.731 & N & Y \\
COSMOS J100130.38+014304.40 &   150.37659  &  1.71789 & 1.572 & 12.3 & 44.1 & -0.321 & 0.731 & Y & Y\\
\enddata

\tablenotetext{}{Notes.--- Table 2 is published in its entirety in the online edition of the article. A portion is shown here for guidance regarding its form and content.}
\end{deluxetable}

\begin{table}
   \tablenum{1}
\caption{Fraction of Objects Identified as BPT AGN\tablenotemark{a}}
\begin{center}
\begin{tabular}{l|ccc|ccc|ccc}

\hline
$\rm log(L_{\rm IR}/L_{\odot})$ & 	\multicolumn{3}{|c|}{Low Redshift} &	\multicolumn{3}{|c|}{Intermediate Redshift}	& \multicolumn{3}{|c}{High redshift}	   	\\
							&	$<z>$ & \# in bin & \% AGN	&	$<z>$ & \# in bin & \% AGN	&	$<z>$ & \# in bin & \% AGN	\\
\hline
$<10$ 		& 0.07	& 18	& 44	& \nodata	& 0		& \nodata	& \nodata	&	0	& \nodata	\\
$10-11$ 	& 0.11	& 50	& 56	& 0.63		& 1		& 0			& \nodata	&	0	& \nodata	  \\
$11-12$ 	& 0.17	& 9		& 33	& 0.73		& 47	& 74		& 1.3		&	7	& 71		 \\
$>12$ 		& 0.07	& 1		& 0		& 0.91		& 6		& 33		& 1.4		& 53	& 72	 \\

\hline

IRAC AGN 	&\nodata	& \nodata	& \nodata	& 0.70		& 10	& 100		& 1.45		& 33	& 79	 \\
not IRAC AGN	& \nodata	& \nodata	& \nodata			& 0.77		& 49	& 57		& 1.41		& 31	& 64  \\
\hline
X-ray Sources\\
\hline
$\rm log(L_{\rm X})<42$ 		& 0.10	& 13	& 77			& \nodata	& 0		& \nodata	& \nodata	&	0	& \nodata	\\
$\rm log(L_{\rm X})=42-43$ 	& 0.2	& 2	 	& 50    			& 0.73		& 3		& 100		& 1.3		& 2	 	& 50 \\
$\rm log(L_{\rm X})=43-44$ 	& \nodata	& 0		& \nodata	& 0.70		& 9		& 78		& 1.67		&18		& 72		 \\
$\rm log(L_{\rm X})>44$ 		& \nodata	& 0		& \nodata	& 0.70		& 6		& 67		& 1.67		& 23	& 87	 \\

\hline
X-ray detected &0.10	& 15	& 73	& 0.71		& 23	& 74		& 1.46		& 44	& 77	 \\
not X-ray detected	& 0.10	& 68	& 44	& 0.83		& 36	& 64		& 1.41		& 20	& 60  \\
\hline
\end{tabular}
\tablenotetext{0}{$^{a}$BPT AGN are objects that lie above the redshift dependent AGN classification like of Kewley \etal 2013b as shown in Figure 2.}

\end{center}
\end{table}

\end{document}